\documentclass[hidelinks, 10pt, conference, compsocconf]{IEEEtran}
\usepackage{times} 
\usepackage{amsmath} 
\usepackage{amssymb}  

\usepackage{hyperref}
\hypersetup{
    colorlinks=true,
    linkcolor=black,
    citecolor=black,
    filecolor=black,
    urlcolor=black,
}

\usepackage{algorithm}
\usepackage{algpseudocode}

\usepackage{graphicx}
\usepackage[usenames,dvipsnames,svgnames,table]{xcolor}
\definecolor{yellow}{HTML}{E2C800}
\definecolor{violet}{HTML}{C136C1}
\usepackage[dot]{dot2texi}
\usepackage[edges]{forest}
\usetikzlibrary{shapes, shapes.geometric, arrows, arrows.meta, matrix, positioning}

\usepackage{cite}

\begin{document}
%
\title{Whole Genome Phylogenetic Tree Reconstruction Using Colored de Bruijn Graphs}




%
\author{\IEEEauthorblockN{Cole A. Lyman\IEEEauthorrefmark{1}, 
M. Stanley Fujimoto\IEEEauthorrefmark{1}, 
Anton Suvorov\IEEEauthorrefmark{2},
Paul M. Bodily\IEEEauthorrefmark{1},\\
Quinn Snell\IEEEauthorrefmark{1},
Keith A. Crandall\IEEEauthorrefmark{3},
Seth M. Bybee\IEEEauthorrefmark{2} and 
Mark J. Clement\IEEEauthorrefmark{1}}
\IEEEauthorblockA{\IEEEauthorrefmark{1}Computer Science Department\\
Brigham Young University,\\
Provo, Utah 84602 USA}
\IEEEauthorblockA{\IEEEauthorrefmark{2}Department of Biology\\
Brigham Young University,\\
Provo, Utah 84602 USA}
\IEEEauthorblockA{\IEEEauthorrefmark{3}Computational Biology Institute\\
George Washington University,\\
Washington, DC 20052 USA}
Email: \href{mailto:colelyman@byu.edu}{colelyman@byu.edu}}


\maketitle

\begin{abstract}
We present \texttt{kleuren}, a novel assembly-free method to reconstruct phylogenetic trees using the Colored de Bruijn Graph.
\texttt{kleuren} works by constructing the Colored de Bruijn Graph and then traversing it, finding bubble structures in the graph that provide phylogenetic signal.
The bubbles are then aligned and concatenated to form a supermatrix, from which a phylogenetic tree is inferred.
We introduce the algorithms that \texttt{kleuren} uses to accomplish this task, and show its performance on reconstructing the phylogenetic tree of 12 \textit{Drosophila} species.
\texttt{kleuren} reconstructed the established phylogenetic tree accurately and is a viable tool for phylogenetic tree reconstruction using whole genome sequences.
Software package available at: \href{https://github.com/Colelyman/kleuren}{https://github.com/Colelyman/kleuren}.
\end{abstract}


%
\IEEEpeerreviewmaketitle

\section{Introduction}

Whole genome sequences are readily available and affordable like never before \cite{NGS} due to the advent of high-throughput Next Generation Sequencing (NGS) which has provided researchers with vast amounts of genomic sequencing data that has transformed the landscape of understanding of genomes.
The field of phylogenetics, which discovers the evolutionary relationship between taxa, has been no exception to this transformation.
Phylogenetics has responded to the copious amounts of high throughput data with novel alignment-free and assembly-free methods \cite{AAF,CVTree} that are better suited \cite{NextGenPhylo} to handle the large amounts of data more efficiently than the traditional alignment-based phylogenetic methods.
The traditional approach to phylogenetic tree reconstruction requires a homology search throughout the genomes of the taxa, a Multiple Sequence Alignment (MSA) of the homologs, and a tree construction from the resulting matrix.
Each of these steps can be computationally expensive and may introduce many unnecessary assumptions that can be avoided by using an alignment-free and assembly-free method.

Alignment-free and assembly-free methods \cite{AlignFreeReview,AlignFreeReview2,NoMSA,Cophylog} don't come without their disadvantages, one of which being that many of these methods abstract away the source of the phylogenetic signal to a method akin to shared kmer-counting.
We propose an assembly-free whole genome phylogenetic tree reconstruction method using the Colored de Bruijn Graph (CdBG) \cite{CdBG}, a data structure that is commonly used for detecting variation and comparing genomes.

The CdBG is similar to a traditional de Bruijn Graph (dBG) in that the substrings of a certain length, referred to as kmers, of a sequence represent the vertices of the dBG and an edge exists between two vertices if the suffix of the first vertex is the prefix of the second vertex.
The CdBG differs from the traditional dBG in that each vertex is associated to an unique color (or set of colors) which could be a differing sample, species, or taxon.

We introduce the \texttt{kleuren} (Dutch for "colors" in tribute of Nicolaas Govert de Bruijn, the de Bruijn graph's namesake) software package which implements our methods.
\texttt{kleuren} works by finding \textit{bubble} regions \cite{CdBG,Bubbles} of the CdBG, which are where one or more colors diverge at a node, which act as pseudo-homologous regions between the taxa. 
The sequence for each taxon in each bubble is then extracted and a MSA is performed, then the MSA's from each bubble are concatenated to form a supermatrix in which a phylogenetic tree of evolution is constructed.

\section{Methods}

\subsection{Definitions}

Given the alphabet $\Sigma = \{A, C, G, T\}$ which are nucleotide codes, let a dBG $\mathbf{G}$, be defined as $\mathbf{G} = (V, E)$ where $V = \{v_1, v_2, \ldots, v_i, \ldots, v_s\}$ is the set of vertices and where $v_i$ is the $i^{th}$ unique sequence of length $k$ of $\mathbf{G}$ and where $E = \{e_1, e_2, \ldots, e_i, \ldots, e_t\}$ is the set of edges and where $e_i = \left(v_i, v_{i+1}\right)$ is an edge connecting two vertices such that the sequence of $v_i$ and $v_{i+1}$ overlap by $(k-1)$ characters.
Let a CdBG, $\mathbf{CG}$, be defined as $\mathbf{CG} = \{G_1, G_2, \ldots, G_i, \ldots, G_u\}$ for $u$ taxa where $G_i = \left(V_i, E_i\right)$ is the dBG of the $i^{th}$ taxon.
We refer to each $G \in \mathbf{CG}$ as a distinct color or taxon.

\begin{figure}
\centering

\includegraphics[scale=0.55]{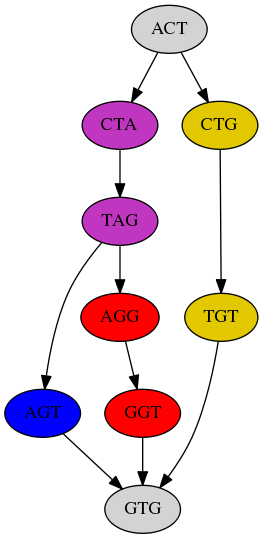}

\begin{flushleft}
\textbf{A.} Bubble in a Colored de Bruijn Graph \\
\end{flushleft}

\begin{tabular}{cl}
{\color{yellow}Color 1} & Path: ACTGTG \\
{\color{red}Color 2} & Path: ACTAGGTG \\ 
{\color{blue}Color 3} & Path: ACTAGTG \\
\end{tabular}

\begin{flushleft}
\textbf{B.} Paths in the Bubble of Each Color \\
\end{flushleft}

\caption{\textbf{A.} An example of a bubble in a Colored de Bruijn Graph with $3$ colors (i.e. $3$ taxa), and where $k=3$.
The colors of the vertices represent the following: gray- all colors contain the vertex, purple- Color 2 and Color 3 contain the vertex, yellow- Color 1 contains the vertex, red- Color 2 contains the vertex, and blue- Color 3 contains the vertex.
In this example \textit{ACT} is the \textit{startVertex} and \textit{GTG} is the \textit{endVertex} which are both contained in all of the colors.
\textbf{B.} The extended paths of each color between the \textit{startVertex} and \textit{endVertex}.
\label{fig:bubble}}
\end{figure}

Furthermore, let a \textit{path}, $P = \left(v_1,\ldots, v_w\right)$ in $G_i$ be defined as a sequence of vertices from $V_i$ such that for all subsequences $\left(v_j,v_{j+1}\right)$ of $P$, the edge $\left(v_j, v_{j+1}\right) \in E_i$. 
Let a \textit{bubble}, $B$, in $\mathbf{CG}$ be defined as $B = \{P_1, \ldots, P_z\}$ such that each $P \in B$ is associated with one or more colors, that the first and last vertices of $\forall P \in B$ are identical, and that $2 \leq z \leq u$ (see Figure~\ref{fig:bubble}).

Finally, let $\mathbf{K}$ be defined as $\mathbf{K} = \{V_1 \cup V_2 \cup \ldots \cup V_i \cup \ldots \cup V_u\}$ where $V_i$ is the vertices (or the unique kmers) of the $i^{th}$ dBG, $\mathbf{G_i}$.

\subsection{Software Architecture}

We use the \texttt{dbgfm} software package \cite{dbgfm} to construct and represent the dBG's of the individual taxa.
\texttt{kleuren} provides an interface to interact with the individual dBG's to create a CdBG, where each taxon is considered a color.
The \texttt{dbgfm} package uses the FM-Index \cite{FM-Index}, as a space efficient representation of the dBG.

\subsection{\texttt{kleuren} Algorithms}

\subsubsection{Overall Algorithm}\label{overall}

\begin{algorithm}
\caption{kleuren Algorithm}\label{overallAlg}
\begin{algorithmic}[1]
	\Function{kleuren}{$\mathbf{K}, \mathbf{CG}$}
    	\State $bubbles \gets \left[ \ \right]$ \Comment{$bubbles$ is initialized to an empty list}
    	\ForAll{$k \in \mathbf{K}$} 
        	\If{$k$ is in $c$ or more colors of $\mathbf{CG}$}
            	\State $endVertex \gets$ \Call{findEndVertex}{$k, \mathbf{CG}$}
                \ForAll{$color \in \mathbf{CG}$}
                	\State $path \gets$ \Call{extendPath}{$k, endVertex,$ $color$}
                    \State add $path$ to $bubble$
                \EndFor
                \State append $bubble$ to $bubbles$
            \EndIf
        \EndFor
        \State $alignments \gets \left[ \ \right]$  
        \ForAll{$bubble \in bubbles$}
        	\State $alignment \gets$ multiple sequence alignment of each $path$ in $bubble$
            \State append $alignment$ to $alignments$
        \EndFor
        \State $supermatrix \gets$ concatenation of $alignments$
    \EndFunction
\end{algorithmic}
\end{algorithm}

\texttt{kleuren} works by iterating over the superset of vertices, $\mathbf{K}$, and discovering vertices that could form a \textit{bubble}.
A vertex, $s$, could form a \textit{bubble} if $s$ is present in $c$ or more colors of $\mathbf{CG}$, where $c$ is set by the user as a command line parameter.
Note that the lower that $c$ is, the more potential bubbles that may be found, but \texttt{kleuren} will take longer to run because more vertices will be considered as the starting vertex of a \textit{bubble}.
Let $s$ be considered as the starting vertex of the \textit{bubble}, $b$; then the end vertex, $e$, of $b$ is found (see Section~\ref{findEndVertex}).
After the end vertex is found, the path, $p$, between $s$ and $e$ is found for each color in $\mathbf{CG}$ (see Section~\ref{extendPath}).
This process is repeated until each vertex in $\mathbf{K}$ has been either considered as a starting vertex of a \textit{bubble}, or has been visited while extending the path between a starting and ending vertex.

\subsubsection{Finding the End Vertex}\label{findEndVertex}

\begin{algorithm}
\caption{Find End Vertex Function}\label{findEndVertexAlg}
\begin{algorithmic}[1]
	\Function{findEndVertex}{$startVertex, \mathbf{CG}$}
    	\State $endVertex \gets ``~"$ \Comment{$endVertex$ is initialized to an empty string}
        \State $neighbors \gets \Call{getNeighbors}{startVertex}$
    	\While{$!\Call{isEmpty}{neighbors}\ and\ $ $\Call{isEmpty}{endVertex}$}
        	\ForAll{$neighbor \in neighbors$}
            	\If{$k$ is in $c$ or more colors of $\mathbf{CG}$}
                	\State $endVertex \gets neighbor$
                \EndIf
            \EndFor
        \EndWhile
        \State \Return{$endVertex$}
    \EndFunction
\end{algorithmic}
\end{algorithm}

The end vertex is found by traversing the path from the $startVertex$ until a vertex is found that is in at least $c$ colors.
The $endVertex$ is then used in the function to extend the path (see Section~\ref{extendPath}). 

\subsubsection{Extending the Path}\label{extendPath}

\begin{algorithm}
\caption{Extend the Path Functions}\label{extendPathAlg}
\begin{algorithmic}[1]
	\Function{extendPath}{$startVertex, endVertex, color$, $maxDepth$}
    	\State $path \gets ``~"$
        \State $visited \gets \{\}$ \Comment{$visited$ is initialized to the empty set}
    	\If{\Call{recursivePath}{$startVertex, endVertex, path,$ $color,visited, 0, maxDepth$}}
        	\State \Return{$path$}
        \EndIf
    \EndFunction
    \\
    \Function{recursivePath}{$currentVertex, endVertex,$ $path, color, visited, depth, maxDepth$}
    	\State add $currentVertex$ to $visited$
        \If{$depth >= maxDepth$}
        	\State \Return $false$
        \EndIf
        \If{$currentKmer == endKmer$}
        	\State \Return $true$
        \EndIf
        \State $neighbors \gets \Call{neighbors}{currentVertex, color}$
        \ForAll{$neighbor \in neighbors$}
        	\If{$neighbor$ is in $visited$}
            	\State continue
            \EndIf
            \State $oldPath \gets path$ 
            \State append suffix of $currentKmer$ to $path$
            \State $depth \gets depth + 1$
            \If{!\Call{recursivePath}{$neighbor, endVertex, path,$ $color, visited, depth, maxDepth$}}
            	\State $path \gets oldPath$
            \Else
            	\State \Return{$true$}
            \EndIf
        \EndFor
    \EndFunction
\end{algorithmic}
\end{algorithm}

The main functions that discover the sequences found in a bubble are the Extend the Path Functions (see Section~\ref{extendPath}).
To extend the $path$ between the $startVertex$ and $endVertex$ we use a recursive function that traverses the dBG for a color in which every possible path between the $startVertex$ and $endVertex$ is explored up to the $maxDepth$ (provided as a command line parameter by the user).
The $maxDepth$ parameter allows the user to specify how thorough \texttt{kleuren} will search for a \textit{bubble}; the higher the $maxDepth$ the more \textit{bubbles} that \texttt{kleuren} will potentially find, but the longer \texttt{kleuren} will take because at each depth there are exponentially more potential paths to traverse.

\subsection{Data Acquisition}

To measure the effectiveness of our method we used 12 \textit{Drosophila} species, obtained from FlyBase \cite{FlyBase}.
We chose this group of species because there is a thoroughly researched and established phylogenetic tree \cite{Hahn-true-tree}.

\subsection{Tree Construction and Parameters}\label{sec:tree-construction}

We used the DSK software package \cite{DSK} to count the kmers present in all of the \textit{Drosophila} species.
To find the bubbles, we used the following parameters: $k = 17$ (kmer size of $17$) and $c = 12$ (all colors in the $\mathbf{CG}$ were required to contain a vertex in order to search for a bubble starting at that vertex) and ran $32$ instances of \texttt{kleuren} concurrently for $4$ days to find $3,277$ bubbles.
When all of the \textit{bubbles} in the CdBG had been identified, we used MAFFT \cite{MAFFT} to perform a MSA for each sequence in every \textit{bubble} that \texttt{kleuren} identified (see Figure~\ref{fig:overall} A.).
Then each MSA was concatenated to form a supermatrix (see Figure~\ref{fig:overall} B.) using Biopython \cite{Biopython}.
The phylogenetic tree was then inferred from the supermatrix by Maximum Likelihood using IQ-TREE \cite{iqtree} (see Figure~\ref{fig:overall} C.).

Once the tree was constructed, we used the ETE 3 software package \cite{ETE3} to compare the tree to the established one and Phylo.io \cite{phylo.io} to visualize the trees.

\subsection{Bubble Assumptions}

Our method is based on the assumption that \textit{bubbles} are representative of homologous regions of the taxa genomes.
We propose that this assumption is reliable because it has been shown that dBG's are a suitable method to align sequences \cite{MultipleAlignment,Sibelia,TwoPaCo}, and by identifying the \textit{bubbles} in the CdBG we find the sections of the graph that contain the most phylogenetic signal.

\begin{figure}
\centering

\begin{tabular}{cl}
{\color{yellow}Color 1} & Path: ACT\texttt{-}\texttt{-}GTG \\
{\color{red}Color 2} & Path: ACTAGGTG \\ 
{\color{blue}Color 3} & Path: ACTA\texttt{-}GTG \\
\end{tabular}

\begin{flushleft}
\textbf{A.} Multiple Sequence Alignment of the Sequences in Bubble (Figure~\ref{fig:bubble}) \\
\end{flushleft}

\medskip

\begin{tabular}{cl}
{\color{yellow}Color 1} & Path: ACT\texttt{-}\texttt{-}GTGATT\texttt{-}A... \\
{\color{red}Color 2} & Path: ACTAGGTGATTC\texttt{-}... \\ 
{\color{blue}Color 3} & Path: ACTA\texttt{-}GTGATTCA... \\
\end{tabular}

\begin{flushleft}
\textbf{B.} Supermatrix of Multiple Sequence Alignments concatenated \\
\end{flushleft}

\medskip

\begin{forest}
  forked edges,
  /tikz/every pin edge/.append style={Latex-, shorten <=2.5pt, darkgray},
  /tikz/every pin/.append style={darkgray},
  /tikz/every label/.append style={darkgray},
  before typesetting nodes={
    delay={
      where content={}{coordinate}{},
    },
    where n children=0{tier=terminus, label/.wrap pgfmath arg={right:#1}{content()}, content=}{},
  },
  for tree={
    grow'=0,
    s sep'+=10pt,
    l sep'+=15pt,
  },
  l sep'+=10pt,
  [, 
      [
        [\color{blue} Color 3]
        [\color{red} Color 2]
      ]
      [\color{yellow} Color 1]
  ]
\end{forest}

\begin{flushleft}
\textbf{C.} Phylogenetic Tree \\
\end{flushleft}

\caption{\textbf{A.} The Multiple Sequence Alignment (MSA) of the sequences from the bubble presented in Figure~\ref{fig:bubble}.
\textbf{B.} The MSA's from each bubble are concatenated into a supermatrix, from which a phylogenetic tree is constructed.
\textbf{C.} The resulting tree from the supermatrix inferred by Maximum Likelihood.
\label{fig:overall}}
\end{figure}

\begin{figure*}
\centering
\includegraphics[scale=0.45]{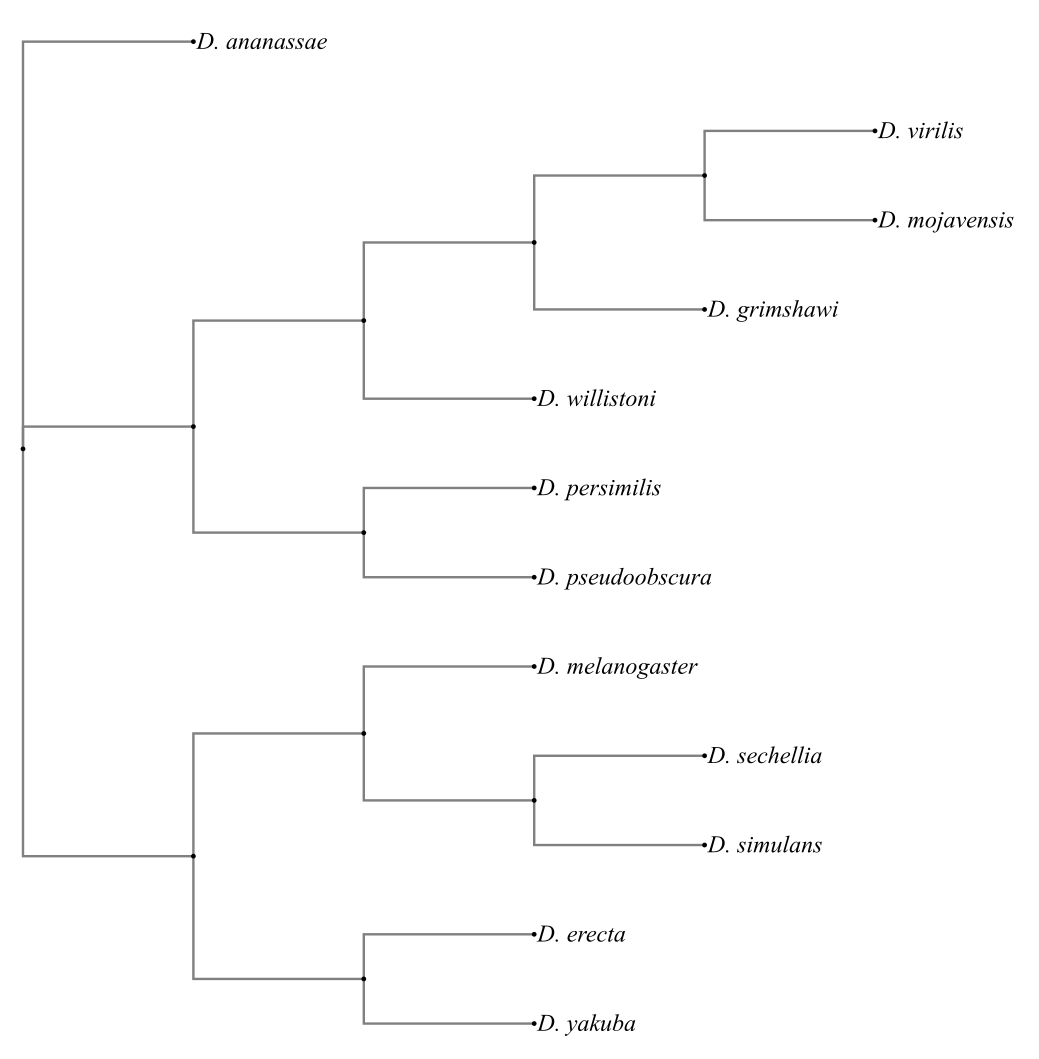}

\caption{The phylogenetic tree of 12 \textit{Drosophila} species constructed using \texttt{kleuren}. 
This tree resulted from using a kmer size of 17 and required all species to contain a vertex in order for the algorithm to search for a bubble starting at that vertex; and this tree is consistent with the established tree for these 12 species.
\label{fig:tree}}
\end{figure*}

\section{Results}\label{results}

\texttt{kleuren} constructed a tree (see Figure~\ref{fig:tree}) consistent with the established tree found in \cite{Hahn-true-tree} (the Robinson-Foulds distance \cite{Robinson1981} between the two trees is 0).
Even though we ran many concurrent instances of \texttt{kleuren} for multiple days (see Section~\ref{sec:tree-construction}), not all of the kmers in $\mathbf{K}$ were explored for potential bubbles; meaning that many more bubbles could be found in this CdBG which would only make the phylogeny more concrete.

Before this final successful run, there were a number of unsuccessful attempts made to construct the tree.
Initial attempts were unsuccessful because $\mathbf{K}$ (the super-set of kmers) that \texttt{kleuren} uses to find bubbles was semi-sorted (segments of the file were sorted, but all of the kmers in the file were not in lexicographic order) so the vertices that \texttt{kleuren} used to search for bubbles were skewed towards vertices that were lexicographically first.
We remedied this issue by shuffling the order of the kmer file so that there was no lexicographic bias towards the bubbles that \texttt{kleuren} finds.

A previous attempt resulted in a tree that had a $0.44$ normalized Robinson-Fould's distance from the established tree occurred because there were too few bubbles, and therefore there was not enough phylogenetic signal for the correct tree to be constructed.
To find more bubbles, we split up the kmer file into parts so that multiple instances of \texttt{kleuren} could find bubbles concurrently.
We also discovered that there was a high frequency of adenines (A) (a frequency around $40\%$ in comparison to the other nucleotides) in the final supermatrix that could skew the final tree because nucleotides have differing mutation rates.
We thought this bias towards A was due to the fact that in the $recursivePath$ function (see Algorithm~\ref{extendPathAlg}) the $neighbors$ may be sorted, so the function would traverse the $neighbor$ that started with an A before traversing the other $neighbors$ (see Algorithm~\ref{extendPathAlg}, line: 18).
Similar to the previous sorting problem, we shuffled the order of the $neighbors$ so that the first $neighbor$ that was traversed would not always be lexicographically first.
Despite this change, the final supermatrix that produced the true tree still had a bias towards A (see Section~\ref{futureWork}).

\section{Conclusion}

We introduced a novel method of constructing accurate phylogenetic trees using a CdBG.
Our method, \texttt{kleuren}, uses whole genome sequences to construct a CdBG representation, then it traverses the CdBG to discover bubble structures which become the basis for phylogenetic signal between taxa and eventually produces a phylogenetic tree.

As the NGS era progresses, whole genome sequences are becoming more prevalent for more non-model organisms, in which phylogenies of these organisms have never been constructed.
\texttt{kleuren} is a viable method to relatively quickly and accurately construct the phylogenies for these newly sequenced organisms.


\section{Future Work} \label{futureWork}

We plan to optimize \texttt{kleuren} so that it can find more bubbles in a shorter amount of time.
We will do this by replacing the underlying data structure for how the CdBG is represented.
\texttt{dbgfm}, the current method used to represent the dBG in \texttt{kleuren}, sacrifices time efficiency for memory efficiency by storing the FM-Index entirely on disk, thus slowing down queries into the dBG.
When \texttt{kleuren} runs faster, more bubbles will be found, and more phylogenetic signal will be present so that a more accurate tree can be constructed. 

We also plan to investigate the reasons for the high abundance of A's in the supermatrix (see Section~\ref{results}) further, and balance the frequency of nucleotides in the supermatrix.

Furthermore, we would like to look into how \texttt{kleuren} performs when the CdBG is constructed using read sequencing data rather than assembled genomes.

\section*{Acknowledgment}

This work was funded through the Utah NASA Space Grant Consortium and EPSCoR and through the BYU Graduate Research Fellowship.

The authors would like to thank Kristi Bresciano, Michael Cormier, Justin B. Miller, Brandon Pickett, Nathan Schulzke, and Sage Wright for their thoughts concerning the project.
The authors would also like to thank the Fulton Supercomputing Laboratory at Brigham Young University for their work to maintain the super-computer on which these experiments were run.

\newpage



\bibliographystyle{IEEEtran}
\bibliography{biblio}

\begin{thebibliography}{10}
\providecommand{\url}[1]{#1}
\csname url@samestyle\endcsname
\providecommand{\newblock}{\relax}
\providecommand{\bibinfo}[2]{#2}
\providecommand{\BIBentrySTDinterwordspacing}{\spaceskip=0pt\relax}
\providecommand{\BIBentryALTinterwordstretchfactor}{4}
\providecommand{\BIBentryALTinterwordspacing}{\spaceskip=\fontdimen2\font plus
\BIBentryALTinterwordstretchfactor\fontdimen3\font minus
  \fontdimen4\font\relax}
\providecommand{\BIBforeignlanguage}[2]{{%
\expandafter\ifx\csname l@#1\endcsname\relax
\typeout{** WARNING: IEEEtran.bst: No hyphenation pattern has been}%
\typeout{** loaded for the language `#1'. Using the pattern for}%
\typeout{** the default language instead.}%
\else
\language=\csname l@#1\endcsname
\fi
#2}}
\providecommand{\BIBdecl}{\relax}
\BIBdecl

\bibitem{NGS}
\BIBentryALTinterwordspacing
S.~C. Schuster, ``Next-generation sequencing transforms today's biology,''
  \emph{Nature Methods}, vol.~5, no.~1, pp. 16--18, dec 2007. [Online].
  Available: \url{https://doi.org/10.1038/nmeth1156}
\BIBentrySTDinterwordspacing

\bibitem{AAF}
\BIBentryALTinterwordspacing
H.~Fan, A.~R. Ives, Y.~Surget-Groba, and C.~H. Cannon, ``An assembly and
  alignment-free method of phylogeny reconstruction from next-generation
  sequencing data,'' \emph{{BMC} Genomics}, vol.~16, no.~1, jul 2015. [Online].
  Available: \url{https://doi.org/10.1186/s12864-015-1647-5}
\BIBentrySTDinterwordspacing

\bibitem{CVTree}
\BIBentryALTinterwordspacing
J.~Qi, H.~Luo, and B.~Hao, ``{CVTree}: a phylogenetic tree reconstruction tool
  based on whole genomes,'' \emph{Nucleic Acids Research}, vol.~32, no. Web
  Server, pp. W45--W47, jul 2004. [Online]. Available:
  \url{https://doi.org/10.1093/nar/gkh362}
\BIBentrySTDinterwordspacing

\bibitem{NextGenPhylo}
\BIBentryALTinterwordspacing
C.~X. Chan and M.~A. Ragan, ``Next-generation phylogenomics,'' \emph{Biology
  Direct}, vol.~8, no.~1, jan 2013. [Online]. Available:
  \url{https://doi.org/10.1186/1745-6150-8-3}
\BIBentrySTDinterwordspacing

\bibitem{AlignFreeReview}
\BIBentryALTinterwordspacing
B.~Haubold, ``Alignment-free phylogenetics and population genetics,''
  \emph{Briefings in Bioinformatics}, vol.~15, no.~3, pp. 407--418, nov 2013.
  [Online]. Available: \url{https://doi.org/10.1093/bib/bbt083}
\BIBentrySTDinterwordspacing

\bibitem{AlignFreeReview2}
\BIBentryALTinterwordspacing
O.~Bonham-Carter, J.~Steele, and D.~Bastola, ``Alignment-free genetic sequence
  comparisons: a review of recent approaches by word analysis,''
  \emph{Briefings in Bioinformatics}, vol.~15, no.~6, pp. 890--905, jul 2013.
  [Online]. Available: \url{https://doi.org/10.1093/bib/bbt052}
\BIBentrySTDinterwordspacing

\bibitem{NoMSA}
\BIBentryALTinterwordspacing
C.~X. Chan, G.~Bernard, O.~Poirion, J.~M. Hogan, and M.~A. Ragan, ``Inferring
  phylogenies of evolving sequences without multiple sequence alignment,''
  \emph{Scientific Reports}, vol.~4, no.~1, sep 2014. [Online]. Available:
  \url{https://doi.org/10.1038/srep06504}
\BIBentrySTDinterwordspacing

\bibitem{Cophylog}
\BIBentryALTinterwordspacing
H.~Yi and L.~Jin, ``Co-phylog: an assembly-free phylogenomic approach for
  closely related organisms,'' \emph{Nucleic Acids Research}, vol.~41, no.~7,
  pp. e75--e75, jan 2013. [Online]. Available:
  \url{https://doi.org/10.1093/nar/gkt003}
\BIBentrySTDinterwordspacing

\bibitem{CdBG}
\BIBentryALTinterwordspacing
Z.~Iqbal, M.~Caccamo, I.~Turner, P.~Flicek, and G.~McVean, ``De novo assembly
  and genotyping of variants using colored de {B}ruijn graphs,'' \emph{Nature
  Genetics}, vol.~44, no.~2, pp. 226--232, jan 2012. [Online]. Available:
  \url{https://doi.org/10.1038/ng.1028}
\BIBentrySTDinterwordspacing

\bibitem{Bubbles}
\BIBentryALTinterwordspacing
G.~Peng, P.~Ji, and F.~Zhao, ``A novel codon-based de {B}ruijn graph algorithm
  for gene construction from unassembled transcriptomes,'' \emph{Genome
  Biology}, vol.~17, no.~1, nov 2016. [Online]. Available:
  \url{https://doi.org/10.1186/s13059-016-1094-x}
\BIBentrySTDinterwordspacing

\bibitem{dbgfm}
\BIBentryALTinterwordspacing
R.~Chikhi, A.~Limasset, S.~Jackman, J.~T. Simpson, and P.~Medvedev, ``On the
  representation of de bruijn graphs,'' in \emph{Research in Computational
  Molecular Biology: 18th Annual International Conference, RECOMB 2014,
  Pittsburgh, PA, USA, April 2-5, 2014, Proceedings}, R.~Sharan, Ed.\hskip 1em
  plus 0.5em minus 0.4em\relax Cham: Springer International Publishing, 2014,
  pp. 35--55. [Online]. Available:
  \url{https://doi.org/10.1007/978-3-319-05269-4_4}
\BIBentrySTDinterwordspacing

\bibitem{FM-Index}
\BIBentryALTinterwordspacing
P.~Ferragina and G.~Manzini, ``Opportunistic data structures with
  applications,'' in \emph{Proceedings of the 41st Annual Symposium on
  Foundations of Computer Science}, ser. FOCS '00.\hskip 1em plus 0.5em minus
  0.4em\relax Washington, DC, USA: IEEE Computer Society, 2000, pp. 390--.
  [Online]. Available: \url{http://dl.acm.org/citation.cfm?id=795666.796543}
\BIBentrySTDinterwordspacing

\bibitem{FlyBase}
\BIBentryALTinterwordspacing
L.~S. Gramates, S.~J. Marygold, G.~dos Santos, J.-M. Urbano, G.~Antonazzo,
  B.~B. Matthews, A.~J. Rey, C.~J. Tabone, M.~A. Crosby, D.~B. Emmert,
  K.~Falls, J.~L. Goodman, Y.~Hu, L.~Ponting, A.~J. Schroeder, V.~B. Strelets,
  J.~Thurmond, P.~Zhou, and {FlyBase Consortium}, ``{FlyBase} at 25: looking to
  the future,'' \emph{Nucleic Acids Research}, vol.~45, no.~D1, pp. D663--D671,
  oct 2016. [Online]. Available: \url{https://doi.org/10.1093/nar/gkw1016}
\BIBentrySTDinterwordspacing

\bibitem{Hahn-true-tree}
\BIBentryALTinterwordspacing
M.~W. Hahn, M.~V. Han, and S.-G. Han, ``Gene family evolution across 12
  drosophila genomes,'' \emph{{PLoS} Genetics}, vol.~3, no.~11, p. e197, 2007.
  [Online]. Available: \url{https://doi.org/10.1371/journal.pgen.0030197}
\BIBentrySTDinterwordspacing

\bibitem{DSK}
\BIBentryALTinterwordspacing
G.~Rizk, D.~Lavenier, and R.~Chikhi, ``{DSK}: k-mer counting with very low
  memory usage,'' \emph{Bioinformatics}, vol.~29, no.~5, pp. 652--653, jan
  2013. [Online]. Available:
  \url{https://doi.org/10.1093/bioinformatics/btt020}
\BIBentrySTDinterwordspacing

\bibitem{MAFFT}
\BIBentryALTinterwordspacing
K.~Katoh and D.~M. Standley, ``{MAFFT} multiple sequence alignment software
  version 7: Improvements in performance and usability,'' \emph{Molecular
  Biology and Evolution}, vol.~30, no.~4, pp. 772--780, jan 2013. [Online].
  Available: \url{https://doi.org/10.1093/molbev/mst010}
\BIBentrySTDinterwordspacing

\bibitem{Biopython}
\BIBentryALTinterwordspacing
P.~J.~A. Cock, T.~Antao, J.~T. Chang, B.~A. Chapman, C.~J. Cox, A.~Dalke,
  I.~Friedberg, T.~Hamelryck, F.~Kauff, B.~Wilczynski, and M.~J.~L. de~Hoon,
  ``Biopython: freely available {P}ython tools for computational molecular
  biology and bioinformatics,'' \emph{Bioinformatics}, vol.~25, no.~11, pp.
  1422--1423, mar 2009. [Online]. Available:
  \url{https://doi.org/10.1093/bioinformatics/btp163}
\BIBentrySTDinterwordspacing

\bibitem{iqtree}
\BIBentryALTinterwordspacing
L.-T. Nguyen, H.~A. Schmidt, A.~von Haeseler, and B.~Q. Minh, ``{IQ}-{TREE}: A
  fast and effective stochastic algorithm for estimating maximum-likelihood
  phylogenies,'' \emph{Molecular Biology and Evolution}, vol.~32, no.~1, pp.
  268--274, nov 2014. [Online]. Available:
  \url{https://doi.org/10.1093/molbev/msu300}
\BIBentrySTDinterwordspacing

\bibitem{ETE3}
\BIBentryALTinterwordspacing
J.~Huerta-Cepas, F.~Serra, and P.~Bork, ``{ETE} 3: Reconstruction, analysis,
  and visualization of phylogenomic data,'' \emph{Molecular Biology and
  Evolution}, vol.~33, no.~6, pp. 1635--1638, feb 2016. [Online]. Available:
  \url{https://doi.org/10.1093/molbev/msw046}
\BIBentrySTDinterwordspacing

\bibitem{phylo.io}
\BIBentryALTinterwordspacing
O.~Robinson, D.~Dylus, and C.~Dessimoz, ``{P}hylo.io: Interactive viewing and
  comparison of large phylogenetic trees on the web,'' \emph{Molecular Biology
  and Evolution}, vol.~33, no.~8, pp. 2163--2166, apr 2016. [Online].
  Available: \url{https://doi.org/10.1093/molbev/msw080}
\BIBentrySTDinterwordspacing

\bibitem{MultipleAlignment}
\BIBentryALTinterwordspacing
B.~Raphael, ``A novel method for multiple alignment of sequences with repeated
  and shuffled elements,'' \emph{Genome Research}, vol.~14, no.~11, pp.
  2336--2346, nov 2004. [Online]. Available:
  \url{https://doi.org/10.1101/gr.2657504}
\BIBentrySTDinterwordspacing

\bibitem{Sibelia}
\BIBentryALTinterwordspacing
I.~Minkin, A.~Patel, M.~Kolmogorov, N.~Vyahhi, and S.~Pham, ``{Sibelia}: A
  scalable and comprehensive synteny block generation tool for closely related
  microbial genomes,'' in \emph{Algorithms in Bioinformatics: 13th
  International Workshop, WABI 2013, Sophia Antipolis, France, September 2-4,
  2013. Proceedings}.\hskip 1em plus 0.5em minus 0.4em\relax Berlin,
  Heidelberg: Springer Berlin Heidelberg, 2013, pp. 215--229. [Online].
  Available: \url{http://dx.doi.org/10.1007/978-3-642-40453-5\_17}
\BIBentrySTDinterwordspacing

\bibitem{TwoPaCo}
\BIBentryALTinterwordspacing
I.~Minkin, S.~Pham, and P.~Medvedev, ``{TwoPaCo}: an efficient algorithm to
  build the compacted de {B}ruijn graph from many complete genomes,''
  \emph{Bioinformatics}, p. btw609, sep 2016. [Online]. Available:
  \url{https://doi.org/10.1093/bioinformatics/btw609}
\BIBentrySTDinterwordspacing

\bibitem{Robinson1981}
\BIBentryALTinterwordspacing
D.~Robinson and L.~Foulds, ``Comparison of phylogenetic trees,''
  \emph{Mathematical Biosciences}, vol.~53, no. 1-2, pp. 131--147, feb 1981.
  [Online]. Available: \url{https://doi.org/10.1016/0025-5564(81)90043-2}
\BIBentrySTDinterwordspacing

\end{thebibliography}
%



\end{document}